\documentclass[aps,floatfix,superscriptaddress,showpacs]{revtex4}

\usepackage{epsfig}
\usepackage{amsmath,amssymb}
\usepackage{graphicx,psfrag}
\usepackage{dcolumn}
\usepackage{bm}
\usepackage{slashed}
\usepackage{color}
\usepackage{ulem}

\newcommand{\beq}{\begin{equation}}
\newcommand{\eeq}{\end{equation}}
\newcommand{\bea}{\begin{eqnarray}}
\newcommand{\eea}{\end{eqnarray}}

\newcommand{\nn}{\nonumber\\}

\newcommand\fig[1]     {Fig.\,{\ref{#1}}}

\newcommand\app[1]     {Appendix~\ref{#1}}

\def\eq#1{(\ref{#1})}
\def\s0#1#2{\mbox{\small{$ \frac{#1}{#2} $}}}
\def\0#1#2{\frac{#1}{#2}}

\def\mr#1{{\mathrm{#1}}}

\begin{document}

\title{Theory of superlocalized magnetic nanoparticle hyperthermia: 
\\ rotating versus oscillating fields}

\author{Zs. Isz\'aly} 
\affiliation{University of Debrecen, H-4010 Debrecen P.O.Box 105, Hungary}

\author{I. G. M\'ari\'an}
\affiliation{University of Debrecen, H-4010 Debrecen P.O.Box 105, Hungary}
\affiliation{MTA-DE Particle Physics Research Group, P.O.Box 51, H-4001 Debrecen, Hungary}

\author{I. A. Szab\'o}
\affiliation{University of Debrecen, H-4010 Debrecen P.O.Box 105, Hungary}

\author{A. Trombettoni}
\affiliation{Department of Physics, University of Trieste, Strada Costiera 11, I-34151 Trieste, Italy}
\affiliation{CNR-IOM DEMOCRITOS Simulation Center, Via Bonomea 265, I-34136 Trieste, Italy}
\affiliation{SISSA, Sezione di Trieste, Via Bonomea 265, I-34136 Trieste, Italy}

\author{I. N\'andori}
\affiliation{University of Debrecen, H-4010 Debrecen P.O.Box 105, Hungary}
\affiliation{MTA-DE Particle Physics Research Group, P.O.Box 51, H-4001 Debrecen, Hungary}
\affiliation{Atomki, P.O. Box 51, H-4001 Debrecen, Hungary}

\begin{abstract} 
The main idea of magnetic hyperthermia is to increase locally the temperature of the 
human body by means of injected superparamagnetic nanoparticles. They 
absorb energy from a time-dependent external magnetic field and transfer it into their 
environment. In the so-called \textit{superlocalization}, the combination of an applied 
oscillating and a static magnetic field gradient provides even more focused heating
since for large enough static field the dissipation is considerably reduced. Similar effect 
was found in the deterministic study of the rotating field combined with a static field gradient.
Here we study theoretically the influence of thermal effects on superlocalization and on 
heating efficiency. We demonstrate that when time-dependent steady state 
motions of the magnetisation vector are present in the zero temperature limit, then 
deterministic and stochastic results are very similar to each other. We also show
that when steady state motions are absent, the superlocalization is severely reduced by 
thermal effects. Our most important finding is that in the low frequency range ($\omega \to 0$) 
suitable for hyperthermia, the oscillating applied field is shown to result in two times larger 
intrinsic loss power and specific absorption rate then the rotating one with identical 
superlocalization ability which has importance in technical realisation.
\end{abstract}

\maketitle 

%------------------------------------------------------------------------------------ 
\section{Introduction}
\label{sec_intro}
%------------------------------------------------------------------------------------
Magnetic nanoparticle hyperthermia \cite{pankhurst,ortega_d,pankhurst_q_a,perigo_e_a,cabrera_d,clinapp_1,clinapp_2}, 
or magnetic fluid hyperthermia, is an adjuvant cancer therapy where magnetic nanoparticles are injected into the body.
The nanoparticles are used to transfer energy from the applied (usually alternating) magnetic field in order to locally 
increase the temperature, which is used to treat cancer. The idea to use temperature increase of the human body for 
medical applications has an ancient history. However, the whole-body hyperthermia has limitations since it is very 
difficult to increase the core temperature of the human body and if we do so it could have serious drawbacks. Thus, 
one should turn to local heating. The more local the hyperthermia the more effective the treatment is. In general 
targeted magnetic hyperthermia has a great importance \cite{egfr,domenech_m}. Magnetic nanoparticles can be 
used to achieve a very focused, very well localised heating of the human body. Indeed, there is an increasing interest 
on how to improve the efficiency of the heat transfer, see e.g.,~\cite{adv_func_mat,angew_minireview,review_rinaldi}, 
and also on how to keep the injected particles localised, since they can accumulate in non-targeted tissues 
\cite{non_local_heat}. A spatially controlled and image guided nanoscale magnetic hyperthermia have been 
recently tested {\it in vivo} and {\it in vitro} experiments \cite{mpi_test}. This method is based on a combination 
of an applied alternating and a static field gradient which provides a spatially focused heating
\cite{focused_hyperthermia_1,focused_hyperthermia_2,focused_hyperthermia_3,focused_hyperthermia_4,
recent_focused_hyperthermia}. The reason is that for large enough static field the dissipation drops to zero, 
and therefore the temperature increase is observed mostly where the static field vanishes. This is an instance of 
the phenomenon referred to as "superlocalization", i.e., the controllable localization of the temperature increase 
by the usage of combinations of inhomogeneous static and time-dependent magnetic fields. This can also be 
understood by considering the area of the dynamic hysteresis loops. For oscillating applied field, the area of the 
dynamic hysteresis loops becomes narrow if the static field is present, thus, the amount of energy transferred is 
decreased and one expects a bell-shaped curve of the energy transfer as a function of the static field amplitude, 
see, e.g., Fig.~(10.4) of \cite{review_rinaldi} or for more details \cite{MRSh_theory_focused}. The idea of spatially 
focused, i.e., superlocalized selective heating of a desired region provides us a very precise heating mechanism 
which is realised in recent {\it in vivo} experiments \cite{mpi_test}.

A recent proposal 
\cite{stat_rot_fields} discusses the possibility of a superlocalized and enhanced 
magnetic hyperthermia by an appropriately chosen combination of static and 
rotating applied fields where the amplitudes of the (in-plane) static and rotating 
fields fulfil a certain relation. For example, in case of magnetically isotropic 
nanoparticles the amplitudes of the applied magnetic field (static and rotating) 
should be the same. The advantage of the method presented in \cite{stat_rot_fields} 
is the drastic enhancement of the heat transfer which can be very well localised 
by a static field gradient, i.e.,spatially controlled by the ratio of the amplitudes of 
the static and rotating applied fields. Another recent example is Ref. \cite{kim}, where 
the dynamics and the energy dissipation in the presence of a static and rotating fields 
have been studied and enhancement effects are reported, however the frequency
range is too large to be applicable for hyperthermia. Indeed, the applied external 
magnetic field is one of the most easily variable parameters to increase the efficiency 
of heat generation and one finds several attempts in the literature where the case of 
the rotating external magnetic field has been studied, as an example see the selection
\cite{Bertotti,Denisov_1,Denisov_2,Cantillon,Ahsen,Raikher_Stephanov,Denisov_thermal,Denisov_thermal_1,
Lyutyy,Lyutyy_energy,Chen,exp_rot,lyutyy_general,viscous_rotating}. 

If the applied frequency is high enough ($\omega \gtrsim 100$ kHz) and the diameter 
of the nanoparticle is appropriately chosen (typically small enough), the mechanical 
rotation of  the nanoparticles in the surrounding medium is restricted and one can 
estimate the efficiency of the heat transfer by considering the change in the orientation 
of the magnetization vector, which is called the Neel relaxation process, 
\cite{lyutyy_general,neel_brown,viscous_rotating,hergt_dutz_2}.
On the other hand the applied frequency and  the product of the frequency and field 
amplitude cannot be arbitrarily large due to the biological limitation of hyperthermic 
treatment of the human body \cite{hergt_dutz_1,hergt_dutz_2,hergt_dutz_3}. On the 
one hand, in order to minimise eddy currents, the limiting threshold for frequencies is 
beyond several hundred kHz (certainly not more than 1000 kHz). On the other hand,
the product (of the frequency and field strength) is related to the power 
injected, so, it has to be smaller then an upper bound, the so called Hergt-Dutz limit 
\cite{hergt_dutz_1,hergt_dutz_2,hergt_dutz_3}.
General advice is to use a field frequency of several hundred kHz in 
combination with rather low field amplitude (few kA/m) or a relatively high field amplitude 
(a few tens of kA/m) in combination with a frequency of a few hundred kHz. Therefore, 
if the applied frequency and the applied field amplitude are chosen appropriately, it can 
be used for hyperthermia and in addition the energy loss depends on the dynamics of 
the magnetization only, which can be described by the so called Landau-Lifshitz-Gilbert 
(LLG) equation \cite{LLG_1,LLG_2}. Let us note that a different theoretical approach to study 
the relaxation and obtain the dynamic hysteresis loop is the use of the so called 
Martensyuk, Raikher, and Shliomis (MRSh) equation, which has been applied to 
consider the focused (superlocalized) hyperthermia in \cite{MRSh_theory_focused}. 
The MRSh equation stands for the average magnetisation but the LLG equation 
describes the dynamics of a single magnetic moment which is used to calculate the 
time-dependence of the average moment. In this work we rely on the LLG approach.

In view of experimental realisations, it is crucial to consider the influence of thermal 
fluctuations. Such effects were not taken into account in study of the dynamics of the 
magnetization performed using the deterministic LLG equation, such as in \cite{stat_rot_fields}.
The effect of thermal fluctuations can be studied by using the stochastic LLG approach 
\cite{sLLG_1,sLLG_2}. The motivation of the present work is to explore how thermal fluctuations can 
affect the superlocalization in presence of static and alternating (oscillating or rotating) fields. 
Thus, our main goal here is to investigate the influence of thermal effects by means of the 
stochastic LLG equation on superlocalization and heating efficiency, and to compare 
the rotating and oscillating cases in the presence of a static magnetic field. 

Although, we show that the combination of static and rotating fields can have importance 
in the technical realisation of magnetic nanoparticle hyperthermia, we put serious limitations 
on the heating efficiency of the rotating applied field compared to the usual oscillating one.

%--------------------------------------------------------------------------------------
\section{Deterministic and stochastic LLG equations}
\label{sec_llg}
%--------------------------------------------------------------------------------------
In the following, we use various types of approximations which simplifies the treatment 
and allow for to get qualitative results on the dependence of the heat transfer on the 
parameters of the magnetic fields. First, we shall consider individual nanoparticles 
with a single magnetic domain, so we do not take into account the possible aggregation 
of these isolated particles, which usually decreases the efficiency of the heat transfer. 
As mentioned above, only the change of the orientation of the magnetic moments are 
considered and we neglect the effects coming from the rotation of the particle as a whole 
which is a suitable approximation if the applied frequency is high and the diameter of the 
particles is small enough. We do not study the dependence of our results on the size of 
the nanoparticles. Therefore, we use a typical choice for their diameter which is $\sim 20$ nm,
see below for more comments on this point. Finally, we assume magnetically isotropic nanoparticles.

As discussed in the introduction, we use here the LLG equation to determine
the motion of the magnetic moment of a single nanoparticle. The Gilbert-form of the 
deterministic LLG equation reads 
\begin{equation}
\label{G}
\frac{\rm{d}}{\rm{d}t} {\bf M} =  
- \mu_0 \gamma_0  {\bf M} \times  \left[{\bf H}_{\mr{eff}} - \eta\frac{\rm{d}}{\rm{d}t} {\bf M}\right],
\end{equation}
with the unit vector ${\bf M} = {\bf m}/m_S$, where {\bf m} stands for the magnetization 
vector of a single-domain particle normalised by the saturation magnetic moment $m_S$
(e.g., $m_S \approx 10^5$ A/m for a single crystal Fe$_3$O$_4$ \cite{Fannin}).
Moreover, $\gamma_0 = 1.76 \times 10^{11}$~Am$^2$/Js is the gyromagnetic ratio, 
$\mu_0 = 4 \pi \times 10^{-7}$~Tm/A (or N/A$^2$) is the permeability of free space, $\eta$ 
is the damping factor and finally ${\bf H}_{\mr{eff}}$ is the local effective 
magnetic field. The LLG equation retains the magnitude of the magnetization (this is why 
one can introduce a unit vector for the magnetization) and only the orientation changes 
in time. It is therefore convenient to use polar coordinates.
\footnote{
Let us note Eq.~\eq{G} is identical to Eq.~(2.1) of \cite{path_int_sllg},
where ${\bf M}$ denotes the un-normalised magnetization
(so, in Ref.~\cite{path_int_sllg} ${\bf M}$ is not a unit vector and
in addition one has to make the following replacement 
$\gamma_0 \to \mu_0 \gamma_0$). Furthermore, Eq.~\eq{G} is 
identical to Eqs.~(1) and (4) of Ref.~\cite{heat_enhance}.}

Following the notation of Refs.~\cite{aniso,Chatel,heat_enhance,stat_rot_fields}, 
the Gilbert-form \eq{G} can be equivalently rewritten as
\begin{equation}
\label{LLG}
\frac{\rm{d}}{\rm{d}t} {\bf M} = -\gamma' [{\bf M \times H_{\rm{eff}}}] 
+ \alpha' [[{\bf M\times H_{\rm{eff}}]\times M}],
\end{equation}
which is the LLG equation where $\gamma' = \mu_0 \gamma_0 /(1+\alpha^2)$, 
$\alpha' = \gamma' \alpha$ with the dimensionless damping 
{\color{blue}$\alpha = \mu_0\gamma_0\eta$}. 
Typical values for the damping parameter $\alpha$ are $\alpha = 0.1$ and $\alpha = 0.3$, 
which have been used in \cite{Lyutyy_energy} and \cite{Giordano}, respectively.
In this paper we use $\alpha = 0.05$, $\alpha = 0.1$ and $\alpha = 0.3$.

The local effective magnetic field contains the external applied and the anisotropy fields. 
As mentioned above, we consider only spherically symmetric nanoparticles without any
crystal field anisotropy so, the nanoparticles are considered as isotropic. Several different 
type of external applied fields combination of static and alternating terms can be considered.
The latter can be either oscillating or rotating:  
\bea
\label{H_rot_def}
\mr{Rotating:} \hskip 0.5cm {\bf H}_{\rm{eff}} = H \, \, \Big(\cos(\omega t) + b_0,\,\,  \sin(\omega t), \,\, 0\Big), \\
\label{H_osc_parallel_def}
\mr{Parallel \,\,  oscillating:} \hskip 0.5cm {\bf H}_{\rm{eff}} = H \, \, \Big(\cos(\omega t) + b_0,\,\,  0, \,\, 0\Big), \\
\label{H_osc_perp_def}
\mr{Perpendicular \,\, oscillating:} \hskip 0.5cm {\bf H}_{\rm{eff}} = H \, \, \Big(\cos(\omega t),\,\,  b_0, \,\, 0\Big),
\eea
where $H$ and $H b_0$ stands for the amplitude of the applied and the static fields, respectively.

The heat transfer is a monotonic increasing function of the product of the applied frequency 
and the field amplitude but at the same time one cannot increase them too much, otherwise 
they would not usable for medical applications \cite{hergt_dutz_1,hergt_dutz_2,hergt_dutz_3}. 
For example, the heating efficiency of magnetic nanoparticles was considered for relatively
high amplitudes $100$ kA/m at $150$ kHz in \cite{high_amplitude}. The upper bound on the
product of amplitude and frequency is the Hergt-Dutz limit and the value $5 \times 10^{8}$ A/(m s) 
is used as a safe operational guideline \cite{review_rinaldi}. In other experimental studies it was 
found that this upper limit might be exceeded by a factor of 10. Indeed, an example for  clinical 
application, one should mention the human-sized magnetic field applicator (developed 
by MagForce Nanotechnologies AG Berlin, Germany) which can generate a $100$ kHz 
alternating magnetic field at a variable field strength of $0 - 18$ kA/m, \cite{magforce_1,magforce_2}
which results in $2 \times 10^{9}$ A/(m s) as an upper limit. To explore the dependence on 
other parameters  it is therefore reasonable to keep the product of the frequency and the 
field amplitude as large as possible. Thus, in this work our choice for the
upper (theoretical) limit is $\omega \approx 1000$ kHz and $H \approx18 $kA/m although 
it is above the safe operational limit. Of course, the Hergt-Dutz limit allow us to use different 
values for $\omega$ and $H$ but always keeping their product below its maximum value 
and in addition keeping the frequency below its maximum which is $\omega \approx 1000$ kHz.
In this work we fix $H = 18$ kA/m but in the very last section where we perform a systematic 
search for the optimal choice for $\omega$ and $H$ both for the oscillating and the rotating 
cases.  We define as well the parameters
\bea
\label{def_L}
\omega_L &=&   H \gamma' \nn
\alpha_N &=& H \alpha'.
\eea
Thus, if one takes $\alpha = 0.1$ (and $H = 18$ kA/m), a good and typical
 choice for parameters suitable for hyperthermia is
  $\omega_L = 4 \times10^9 \, \mr{Hz}$ and
  $\alpha_N =  4 \times 10^8 \, \mr{Hz}$.
  One can introduce the dimensionless parameters 
\bea
\label{dimless_param_def}
\omega &\to&  \omega t_0  \nn
\omega_L =   H \gamma' &\to& \omega_L t_0  \nn
\alpha_N = H \alpha' &\to& \alpha_N t_0.
\eea
where the dimension has been rescaled by a suitably chosen time parameter $t_0$. 
Here we use $t_0 = 0.5 \times 10^{-10}$s. In summary, we use the following sets
of dimensionless parameters:
\bea
\label{dimless_param}
\alpha = 0.05 \, \Rightarrow \, \omega_L \sim  0.2, \,\,\, \alpha_N \sim 0.01,   \nn
\alpha = 0.1 \, \Rightarrow \,\omega_L \sim  0.2, \,\,\, \alpha_N \sim 0.02,   \nn
\alpha = 0.3 \, \Rightarrow \, \omega_L \sim 0.2, \,\,\, \alpha_N \sim 0.06. 
\eea
For hyperthermia one has to take the low-frequency limit, so, a realistic choice for the 
dimensionless frequency is $\omega \sim 10^{-4}$ which corresponds to the 
dimensionful value $\sim 2 \times 10^{6}$ Hz.

As a next step we introduce the stochastic LLG equation \cite{sLLG_1,sLLG_2} where thermal 
fluctuations are taken into account by introducing a random magnetic field, ${\bf H}$, 
in Eq.~\eq{G}:
\begin{equation}
\label{sG}
\frac{\rm{d}}{\rm{d}t} {\bf M} =  
- \mu_0 \gamma_0  {\bf M} \times  \left[({\bf H}_{\mr{eff}} + {\bf H}) - \eta\frac{\rm{d}}{\rm{d}t} {\bf M}\right].
\end{equation}
The corresponding stochastic form of the deterministic LLG equation \eq{LLG} can be written as
\begin{equation}
\label{sLLG}
\frac{\rm{d}}{\rm{d}t} {\bf M} = -\gamma' [{\bf M \times (H_{\rm{eff}}+H)}] 
+ \alpha' [[{\bf M\times (H_{\rm{eff}}+H)]\times M}],
\end{equation}
where the stochastic field, ${\bf H} = (H_x, H_y, H_z)$, consists of Cartesian components which 
are independent Gaussian white noise variables with the following properties,
\begin{equation}
\label{stochastic_field}
\langle H_i(t) \rangle = 0, \hskip 0.5 cm \langle H_i(t_1) H_j(t_2) \rangle = 2 \, D \, \delta_{ij} \, \delta(t_1 -t_2)
\end{equation}
where $i = x, y, z$ and $D$ is a parameter defined using the fluctuation-dissipation theorem
as $D = \eta k_B T/(m_s V \mu_0)$ with the Boltzmann factor $k_B$, absolute temperature $T$ 
and the volume of the particle $V$, see e.g.,~\cite{path_int_sllg}. Let us draw the attention
of the reader to the dependence of the stochastic description on the volume of the particles. 
We do not investigate in particular the dependence of our results on the diameter of the 
individual particles, but the volume cannot be too small since then the nanoparticles will not 
stored in the human body a long enough time and, at the same time it cannot be too large
because then particles have multiple magnetic domains which is not favourable regarding the 
heating efficiency. A typical choice for the diameter is around $20-50$ nm and we adopt the
size $20$nm in this work.

In Eq. (\ref{stochastic_field}) the angular brackets stand for averaging over all possible 
realisations of the stochastic field ${\bf H}(t)$. Due to the Gaussian white noise nature of the 
stochastic field, the random process of ${\bf M}(t)$ is a Markovian one. So, one can use the 
Fokker-Planck formalism. We observe that it is known that in the semi-classical 
limit the Keldysh functional approach reduces to the Fokker-Planck equation. The general path 
integral approach in the framework of effective field theories using renormalization group 
techniques is still an active research field, see e.g.,~\cite{ctp_rg,ctp_rg_1}. 

The stochastic LLG equation \eq{sLLG} can be reduced to (or equivalently rewritten as) a system 
of two stochastic differential equations for the polar $\theta$ and azimuthal $\phi$ angles, see 
Refs.~\cite{Giordano,Denisov_thermal,Denisov_thermal_1}.
In a spherical coordinate system the effect of the thermal bath appears as a Brownian motion on a sphere, which can be described by the Ito process
\begin{align}
\label{spherical_brownian_1}
\frac{\rm{d}}{\rm{d}t} \phi &= 
\frac{1}{\sin\theta} \sqrt{\frac{1}{2\tau_N}} n_\phi,
\\
\label{spherical_brownian_2}
\frac{\rm{d}}{\rm{d}t} \theta &= 
\frac{1}{2\tau_N} \frac{\cos\theta}{\sin\theta} + 
\sqrt{\frac{1}{2\tau_N}} n_\theta.
\end{align}
Here $\tau_N \propto 1/T$ is the N\'eel time for which we used the value $\tau_N=1.5 \times 10^{-7}s$ at $T=300$K.
The role of the random white noise is played by the Gaussian variables $n_\phi$ and $n_\theta$, which satisfy the standard properties 
\begin{equation}
\label{nproperties}
\langle n_\phi(t) \rangle = \langle n_\theta(t) \rangle = 
\langle n_\phi(t_1) n_\theta(t_2) \rangle = 0, \hskip 1cm
\langle n_\phi(t_1) n_\phi(t_2) \rangle = 2 \, \delta(t_1 -t_2), \;\;\;
\langle n_\theta(t_1) n_\theta(t_2) \rangle = 2 \, \delta(t_1 -t_2) .
\end{equation}
Of course these equations are only valid without external fields, however
the rhs. of Eqs.~\eq{spherical_brownian_1} \eq{spherical_brownian_2} represents the only difference between the stochastic and deterministic LLG equation. 
In $T\to 0$ these terms vanish and the stochastic LLG equation reduces to the deterministic one.
We developed codes to solve 
numerically the stochastic differential equations in the presence of a static and rotating 
applied field defined by Eqs. \eq{H_rot_def}, \eq{H_osc_parallel_def} and \eq{H_osc_perp_def}. 
The codes are written in python and Mathematica \cite{wolfram} where the built-in functions for Ito processes are used.
Consistency checks and a specific exapmle are shown in Appendix \ref{sec_checks}. In the more complicated cases below, the procedure is the same, the stochastic LLG equations are solved in the same manner in the spherical coordinate system, however the equations are lengthy without containing meaningful new information and thus they are not written.

The energy loss in a single cycle can be easily obtained if the solution {\bf $\bf M$} of the LLG 
equation is known:
\beq
\label{def_loss}
E = \mu_0 m_S \int_{0}^{\frac{2\pi}{\omega}} \mr{d}t 
\left({\bf H}_{\rm{eff}} \cdot \frac{d{\bf M}}{dt} \right).
\eeq
Two important notes are in order, (i) the above formula contains the average magnetisation 
{\bf $\bf M$}, (ii) it holds for a single particle more precisely it has a dimension of J/m$^3$.

The formula \eq{def_loss} gives exactly the same result as the area of the dynamic hysteresis 
loop which we discuss later. Moreover, the energy loss obtained by either the area of the 
dynamical hysteresis loop or by the energy loss formula \eq{def_loss} is related to the imaginary 
part of the frequency-dependent susceptibility. The product of the applied frequency $f$ and the 
energy $E$ obtained in a single cycle \eq{def_loss} for a single particle and  is related to the 
amount of energy absorbed in a second, i.e, the specific absorption rate (SAR),
\beq
\label{sar}
E f \propto \mr{SAR} = \frac{\Delta T \,\, c}{t}
\eeq
where $\Delta T$ is the temperature increment, $c$ the specific heat and $t$ is the time of the 
heating period. Notice that the SAR is sometimes referred to as the specific loss power.
Furthermore, it is useful to define the so called intrinsic loss power (ILP) \cite{ilp},
\beq
\label{ilp}
\mr{ILP} = \frac{\mr{SAR}}{H^2 f}
\eeq
which can be used to compare experimental (and theoretical) SAR values determined
at different magnetic field frequencies  $f$ and/or amplitudes $H$. Thus, from equation 
\eq{def_loss} we have 
\beq
\frac{E}{2\pi\mu_0 m_s H^2} \propto \mr{ILP}.
\eeq

%--------------------------------------------------------------------------------------
\section{%Deterministic results
{\bf Non-stochastic limit}}
\label{sec_det_res}
%--------------------------------------------------------------------------------------
We consider in this section the non-stochastic, deterministic, zero temperature limit 
of the stochastic LLG.  A feature of the non-stochastic limit, provided by the deterministic 
LLG \eq{LLG} is that  the steady state solutions equation can be obtained. These steady 
state solutions (if they exist) are attractive ones, meaning that, independently of the initial 
conditions for the magnetization vector, the solution of the LLG equation tends to these 
steady state ones. Consequently the energy loss can be calculated by simply inserting 
the steady state solutions into the energy loss formula \eq{def_loss}.
In this section we discuss these steady state solutions for the oscillating and rotating 
cases defined by Eqs. \eq{H_rot_def}, \eq{H_osc_parallel_def}, \eq{H_osc_perp_def}.

%%%%%%%%%%%%%%%%%%%%
\subsection{Rotating case}
\begin{figure}[ht] 
\begin{center} 
\includegraphics[width=7.2cm]{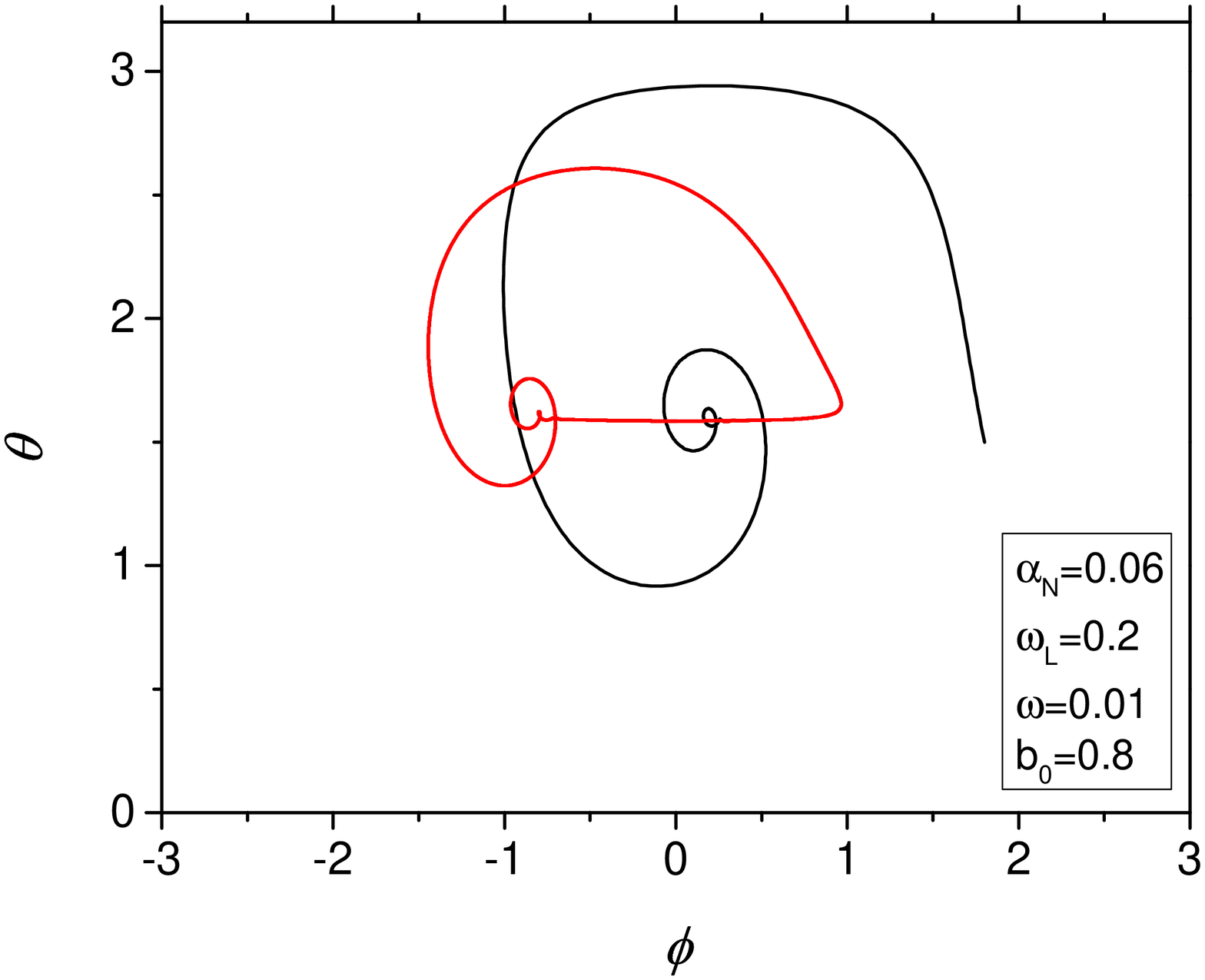}
\caption{(Color online.) Limit-cycle (red line) of the orbit map
in polar coordinates in the rotating frame 
for isotropic nanoparticles when static field is assumed to be in the plane of rotation. 
This result is  obtained by the deterministic LLG equation,
i.e., at zero temperature \cite{stat_rot_fields}, with the 
following dimensionless parameters:
$\alpha_N = 0.06$, $\omega_L = 0.2$, $\omega = 0.01$, and $b_0 = 0.8$.
\label{fig1}
} 
\end{center}
\end{figure}

\begin{figure}[ht] 
\begin{center} 
\includegraphics[width=7.2cm]{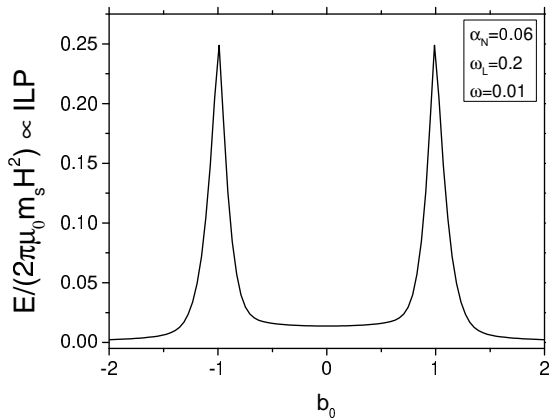}
\caption{Normalised energy loss \eq{def_loss} is plotted against the 
parameter $b_0$ based on the solution of the deterministic LLG equation \eq{LLG},
with parameters $\alpha_N = 0.06$, $\omega_L = 0.2$ and $\omega = 0.01$.
\label{fig2}
} 
\end{center}
\end{figure}

%%%%%%%%%%%%%%%%%%%%
In this subsection we study the  energy losses for the LLG without thermal 
fluctuations in the rotating case:
\beq
\label{H_rot}
{\bf H}_{\rm{eff}} = H \, \, \Big(\cos(\omega t) + b_0,\,\,  \sin(\omega t), \,\, 0\Big).
\eeq
If the static field is in the plane of rotation,
the time-dependent magnetization tends to a steady state
motion which is a limit-cycle in the reference frame attached to 
the rotating field, see \fig{fig1}. 
%
% Figure 1
%
If the static field is {\it perpendicular} to the plane of rotation or if it vanishes, 
the steady state solution of the LLG equation \eq{LLG} is a fixed point in the 
rotating frame. Therefore, instead of red coloured limit cycle of \fig{fig2} one 
finds fixed points, see e.g.,~\cite{aniso,Chatel,heat_enhance,stat_rot_fields}.
It is, however, always true that the steady state solutions are attractive ones 
which means that the magnetization converges very rapidly to these solutions 
which can be used to determine the heat transfer.  

If the amplitude of the static and rotating fields are identical, the energy loss 
obtained by the steady state solution of the deterministic LLG equation has a 
maximum, i.e., for  $\vert b_0 \vert =1$ one finds peaks in \fig{fig2}.
%
% Figure 2
%
The peaks of \fig{fig2} clearly show
that one finds a drastic enhancement of the energy loss for $\vert b_0 \vert =1$. In addition, 
if a  static field gradient has been used than the heat transfer is not just increased, but also 
superlocalized since in this case tissues are heated up significantly only where the amplitudes 
of the static and rotating fields are the same.

Since this result is obtained in the framework of the deterministic LLG equation \eq{LLG},
for experimental realisation one has to take into account thermal fluctuations -- which can be 
done by using the stochastic version of the LLG approach -- to see whether the superlocalization 
is spoiled or not by temperature effects.

%%%%%%%%%%%%%%%%%%%%
\subsection{Vanishing static field, oscillating case}
\begin{figure}[ht] 
\begin{center} 
\includegraphics[width=8.0cm]{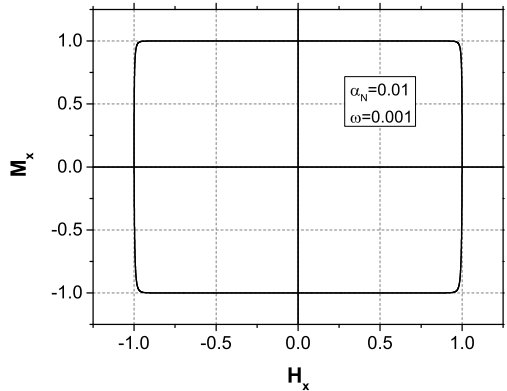}
\caption{Dynamical hysteresis loop based on the solution
\eq{average_sol} of the deterministic LLG equation with the following
parameters, $\alpha_N$ = 0.01, $\omega_L$ = 0.2 and $\omega$ = 0.001.
Its stochastic version is plotted on \fig{fig8}.
\label{fig3}
} 
\end{center}
\end{figure}

%%%%%%%%%%%%%%%%%%%%
Let consider a vanishing static field, i.e., when only oscillating field is applied:
\bea
\label{H_osc}
{\bf H}_{\rm{eff}} = H \, \, \Big(\cos(\omega t),\,\,  0, \,\, 0\Big).
\eea
In this case the solution of the deterministic LLG equation \eq{LLG}
can be obtained analytically \cite{Chatel}:
\begin{eqnarray}
\label{osc_sol_relax}
M_{x}(t) &=& \frac{(M_{x0}-1)+(M_{x0}+1)
\exp{\left[\frac{2\alpha_N}{\omega}\sin(\omega t)\right]}} 
{(1-M_{x0})+(M_{x0}+1)
\exp{\left[\frac{2\alpha_N}{\omega}\sin(\omega t)\right]}},
\nonumber \\
M_{y}(t) &=& \sqrt{1-M_{x}^{2}(t)} \, \,
\sin\left[\frac{\omega_L}{\omega}\sin(\omega t) + \delta_0\right],
\nonumber \\
M_{z}(t) &=& \sqrt{1-M_{x}^{2}(t)} \, \,
\cos\left[\frac{\omega_L}{\omega}\sin(\omega t) + \delta_0\right],
\end{eqnarray}
with the parameters $M_{x0}$ and $\delta_0$ determined by the
initial values $M_{x0} = M_x(0)$ 
and $\delta_0 = \tan^{-1}(M_{y}(0)/M_{z}(0))$. In this particular case,
there are no steady state 
solutions. Thus, \eq{osc_sol_relax} depends on the initial conditions.
The average magnetization reads as
\bea
\langle M_{x}(t) \rangle = \frac{1}{2N}  \sum_{n=1}^{2N} M_{nx}(t) = 
\frac{1}{2N} \sum_{n=1}^{2N} \frac{(M_{nx0}-1)+(M_{nx0}+1)
\exp{\left[\frac{2\alpha_N}{\omega}\sin(\omega t)\right]}} 
{(1-M_{nx0})+(M_{nx0}+1)
\exp{\left[\frac{2\alpha_N}{\omega}\sin(\omega t)\right]}},
\eea
where the summation can be performed pairwise by assuming that for each pair 
$M_{mx0} \approx -M_{mx0}$, which results in
\begin{eqnarray}
\label{average_sol}
\langle M_{x}(t) \rangle &\approx& \frac{1}{N} \sum_{n=1}^{N} \frac{(M_{nx0}^2-1)
\sinh{\left[\frac{2\alpha_N}{\omega}\sin(\omega t)\right]}} 
{-(M_{nx0}^2+1)+(M_{nx0}^2-1)
\cosh{\left[\frac{2\alpha_N}{\omega}\sin(\omega t)\right]}} \nonumber
\\ 
&\approx&  \frac{(\frac{1}{3}-1)
\sinh{\left[\frac{2\alpha_N}{\omega}\sin(\omega t)\right]}} 
{-(\frac{1}{3}+1)+(\frac{1}{3}-1)
\cosh{\left[\frac{2\alpha_N}{\omega}\sin(\omega t)\right]}}
\end{eqnarray}
(as a last step, the summation over $M_{nx0}^2$ is performed assuming a uniform 
distribution). The last expression in \eq{average_sol} can be
used to calculate the energy 
loss in two different ways, either it can be inserted in \eq{def_loss}
directly or one can define a dynamical hysteresis loop
by the substitution $\sin(\omega t) \to \sqrt{1 - H_{\rm{eff},x}^2/H^2}$ 
and calculate the area of the loop. Of course, the two results should be
and indeed are identical. 
For the sake of completeness, a hysteresis loop obtained 
by the solution \eq{average_sol} of the deterministic LLG equation is 
shown in  \fig{fig3}.
%
% Figure 3
%

In order to obtain realistic results one has to use the stochastic LLG equation 
where thermal fluctuations are taken into account. By this modification one can
see whether one retrieves the expected hysteresis loops \cite{usov_hysteresis}.

%%%%%%%%%%%%%%%%%%%%
\subsection{Parallel static field, oscillating case}
%%%%%%%%%%%%%%%%%%%%
Let us now turn to the study of the case where the static field
is chosen to be parallel to the 
direction of oscillation:
\bea
\label{H_osc_parallel}
{\bf H}_{\rm{eff}} = H \, \, \Big(\cos(\omega t) + b_0,\,\,  0, \,\, 0\Big).
\eea
For this configuration of the applied field,
the analytic solution of the deterministic LLG equation is still 
possible and it reads as follows for the x-component
of the magnetization vector
\begin{eqnarray}
\label{osc_sol_relax_para}
M_{x}(t) &=& \frac{(M_{x0}-1)+(M_{x0}+1) \exp{\left[\frac{2\alpha_N}{\omega} (\sin(\omega t) +b_0 \omega t)\right]}} 
{(1-M_{x0})+(M_{x0}+1) \exp{\left[\frac{2\alpha_N}{\omega} (\sin(\omega t) +b_0 \omega t) \right]}}.
\end{eqnarray}
The most important property of the solution \eq{osc_sol_relax_para} is its asymptotic behaviour:
\beq
{\mr{if}} \,\, b_0 > 0, \,\,\, M_x(t\to\infty) = +1, \hskip 1cm
{\mr{if}} \,\, b_0 < 0, \,\,\, M_x(t\to\infty) = -1,
\eeq
which can be seen as a fixed point solution in the laboratory frame,
i.e., it is a time-independent steady state solution.

\begin{figure}[ht] 
\begin{center} 
\includegraphics[width=7.2cm]{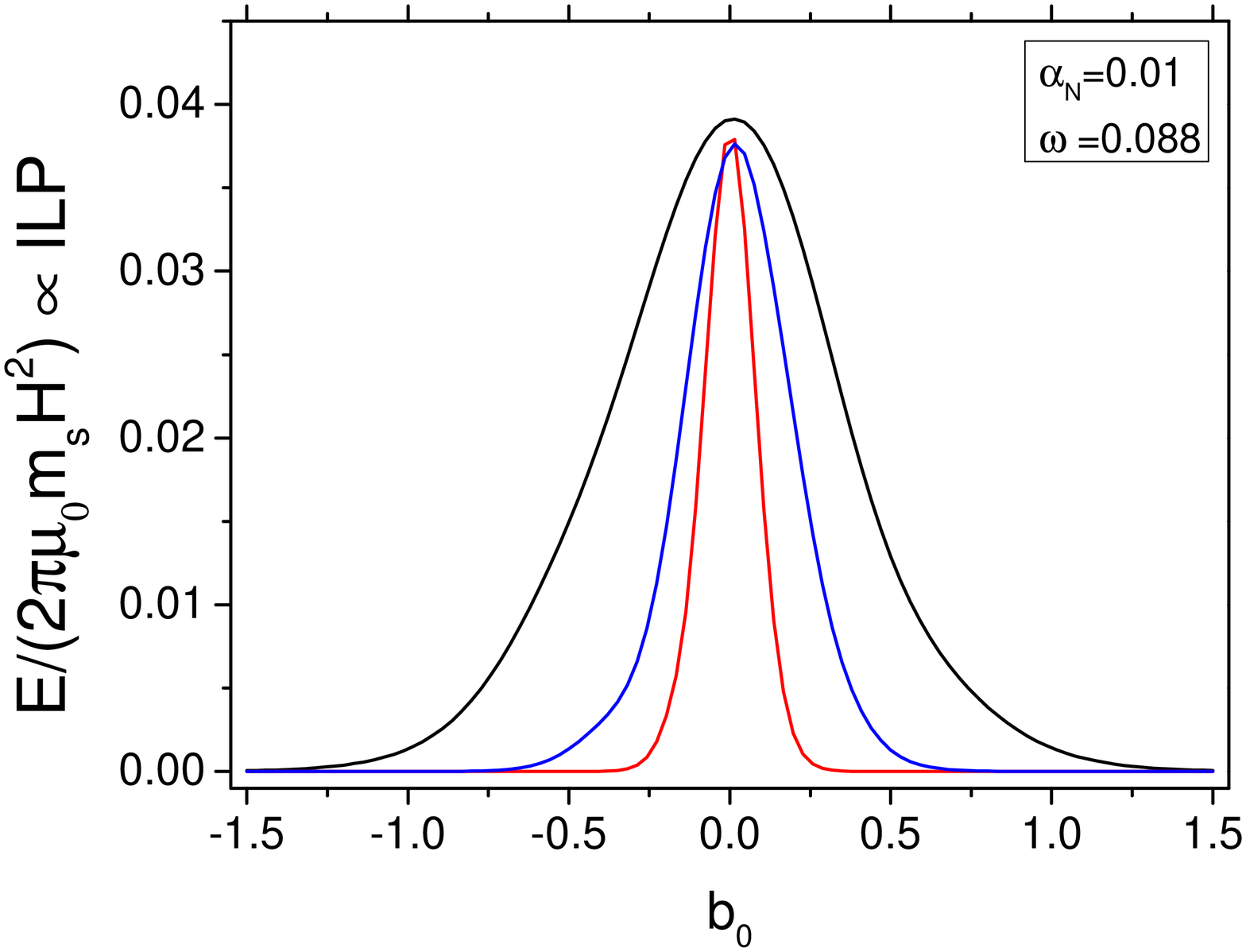}
\caption{(Color online.) Energy loss/cycle as a function of the static field
as based on \eq{osc_sol_relax_para} calculated in different cycles with 
$\alpha_N=0.01$ and $\omega=0.088$, i.e., for $t\to \infty$ the peak 
becomes sharper.
\label{fig4}
} 
\end{center}
\end{figure}

Consequently, the energy loss is basically vanishing according to the 
energy loss formula \eq{def_loss}. In other words,
for $t\to \infty$ the energy loss/cycle is non-zero only 
for $b_0 = 0$, see \fig{fig4}. 
%
% Figure 4
%
This suggests a very sharp superlocalization effect. The combination
of parallel static and oscillating fields 
results in no energy loss for any non-vanishing values of the static field.
However, this picture is based on
the time-independent steady state solution of the deterministic LLG equation
and one may expect a modification due to thermal effects for
small static field values.

%%%%%%%%%%%%%%%%%%%%
\subsection{Perpendicular static field, oscillating case}
%%%%%%%%%%%%%%%%%%%%
Let us now investigate the case where the static field is chosen to
be perpendicular to the direction of oscillation:
\bea
\label{H_osc_perp}
{\bf H}_{\rm{eff}} = H \, \, \Big(\cos(\omega t),\,\,  b_0, \,\, 0\Big).
\eea
For this configuration of the applied field, we rely on numerical solutions.
We observe that for $t\to\infty$ the solution tends to a
time-dependent steady state, which can be seen as a limit cycle
in the laboratory frame. The (time-dependent) steady state solution is
periodic (typically contains higher harmonics) and it
depends on the parameters $\omega, \omega_L, \alpha_N$.
Since it can be seen as an attractive solution, i.e.,
the magnetization vector always tends to this steady state
motion independently of its initial value. 

\begin{figure}[ht] 
\begin{center} 
\includegraphics[width=7.2cm]{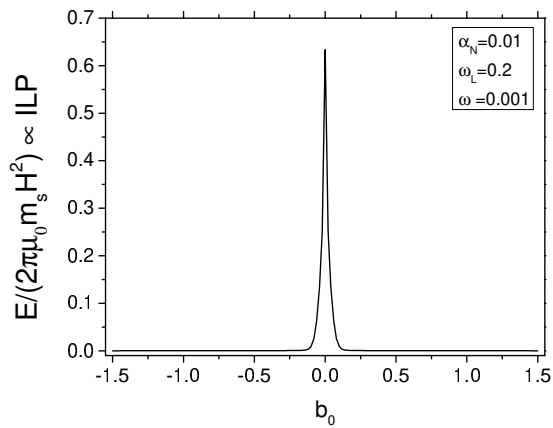}
\caption{Energy loss/cycle as a function of $b_0$ obtained for 
the case where the static field is perpendicular to the line of oscillation with 
parameters $\alpha_N=0.01$, $\omega_L=0.2$ and $\omega=0.001$.
\label{fig5}
} 
\end{center}
\end{figure}

Thus, the energy loss can be calculated by simply taking into account only
this steady state motion. For  the considered values of $\omega, \omega_L, \alpha_N$ 
one finds a single peak in the energy loss at $b_0 = 0$, see \fig{fig5}. This indicates 
that similarly to the previous case, the vanishing static field is favoured, which again 
results in superlocalization.
%
% Figure 5
%

In summary, the solution of the deterministic LLG equation
for the oscillating cases signals a drastic decrease in the energy loss in the 
presence of a non-vanishing static magnetic field independently 
of its orientation. However, this finding needs a careful re-examination 
around the small values of the static field where thermal fluctuations are present, 
since they can have a huge impact on the motion of the magnetization vector. 
Indeed, if one would like to obtain reliable hysteresis loops, the use of the 
stochastic LLG equation is unavoidable. Therefore, we investigate the solution 
of the stochastic LLG equation in the next section.

%--------------------------------------------------------------------------------------
\section{{\bf Stochastic Landau-Lifshitz-Gilbert equation}}
\label{sec_stoc_res}
%--------------------------------------------------------------------------------------
Let us now turn to our main goal and study the stochastic
LLG equation \eq{sLLG} in the presence of the applied magnetic fields
\eq{H_rot_def}, \eq{H_osc_parallel_def} and \eq{H_osc_perp_def},  
which are combinations of static and alternating ones.

%%%%%%%%%%%%%%%%%%%%
\subsection{Rotating case}
\begin{figure}[ht] 
\begin{center} 
\includegraphics[width=7.0cm]{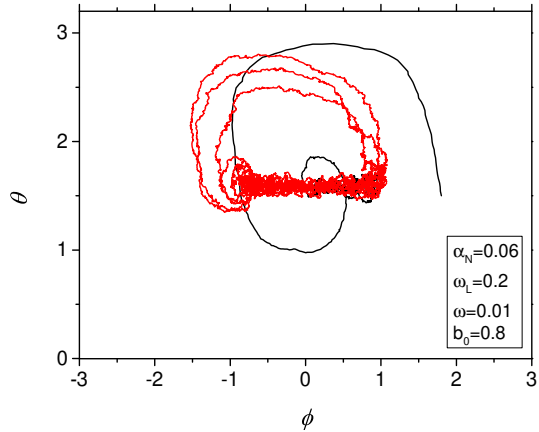}
\caption{(Color online.) The same as \fig{fig1}, but now with non-vanishing 
temperature, i.e., this figure is obtained by the solution of the 
stochastic LLG equation \eq{sLLG} using the parameters and initial values 
of the results shown in \fig{fig1} and $T=300$K.
\label{fig6}
} 
\end{center}
\end{figure}

%%%%%%%%%%%%%%%%%%%%
Let us first consider how the thermal fluctuations modify the steady state
solutions. In \fig{fig6} we show how the limit-cycle of \fig{fig1} obtained by 
the solution of the stochastic LLG equation \eq{sLLG} for the temperature
$T=300$K.
%
% Figure 6
%

\begin{figure}[ht] 
\begin{center} 
\includegraphics[width=7.2cm]{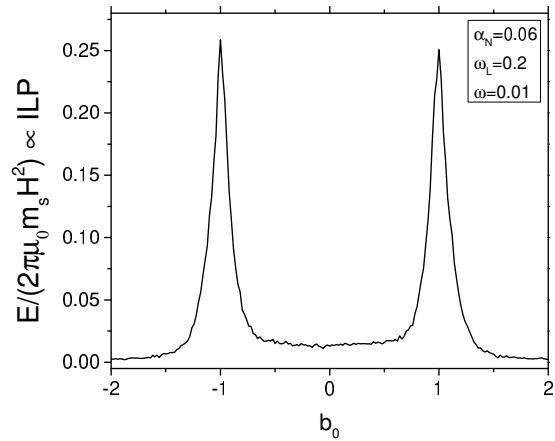}
\caption{The same as \fig{fig2}, but now with non-vanishing temperature ($T=300$K),
i.e., this figure is obtained by the solution of the stochastic LLG equation \eq{sLLG} using 
the parameters of \fig{fig2} and averaged over {\bf $15$} cycles. The energy loss/cycle is plotted 
in terms of the parameter $b_0$.
\label{fig7}
} 
\end{center}
\end{figure}

The next step is to study if and how the thermal fluctuations change the energy
loss obtained for the rotating case. In other words, we aim at studying
the modifications of \fig{fig2} by thermal effects. \fig{fig7} shows the influence of 
thermal fluctuations on the enhancement and superlocalization effects reported 
in the case of the deterministic LLG equation \eq{LLG}.
%
% Figure 7
%

The most important conclusion is that the deterministic (see \fig{fig2})
and the stochastic (see \fig{fig7}) results are very similar to each other.
Thermal effects do not substantially spoil the enhancement and superlocalization 
effect for the rotating case. This is because the energy loss is calculated by using the
(time-dependent) steady state solution of the LLG equations (for the deterministic and 
the stochastic cases) and these steady state solutions are attractive ones, so, the 
magnetization vector always tends to them. It follows that thermal fluctuations are 
not able to significantly modify the dynamics.

%%%%%%%%%%%%%%%%%%%%
\subsection{Vanishing static field, oscillating case}
\begin{figure}[ht] 
\begin{center} 
\includegraphics[width=7.0cm]{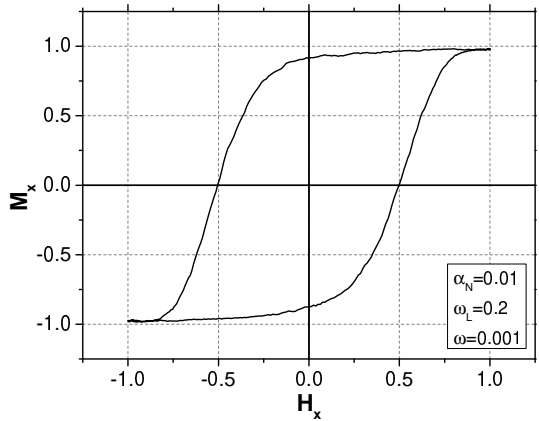}
\caption{The same as \fig{fig3} but for the stochastic version. The hysteresis loop 
is obtained by the stochastic LLG equation averaged over {\bf $50$} runs with $\alpha_N = 0.01$,  
$\omega_L = 0.2$, $\omega = 0.001$ and $T = 300$K. 
\label{fig8}
} 
\end{center}
\end{figure}

%%%%%%%%%%%%%%%%%%%%
Regarding the oscillating case, as a first step, we plot the dynamical hysteresis loop obtained for 
the vanishing static field case, but now with thermal effects taken into account, see \fig{fig8},  
with the parameters $\alpha_N = 0.01, \omega_L = 0.2$, $\omega = 0.001$ and $T = 300$K.
%
% Figure 8
%

This is the expected usual form of the hysteresis loop \cite{usov_hysteresis}, where the
dynamics of the magnetization is averaged out. This represents a drastic modification compared to the 
deterministic hysteresis loop. The reason for such significant change is the absence 
of time-dependent steady state solutions. In other words, the deterministic case is found to be 
an artifact since for any non-vanishing temperature the deterministic picture is spoiled by 
thermal effects. Thus the only way to obtain realistic results for the energy loss when no 
time-dependent steady state motion exits is to use the stochastic LLG equation \eq{sLLG}.

%%%%%%%%%%%%%%%%%%%%
\subsection{Parallel static field, oscillating case}
\begin{figure}[ht] 
\begin{center} 
\includegraphics[width=7.2cm]{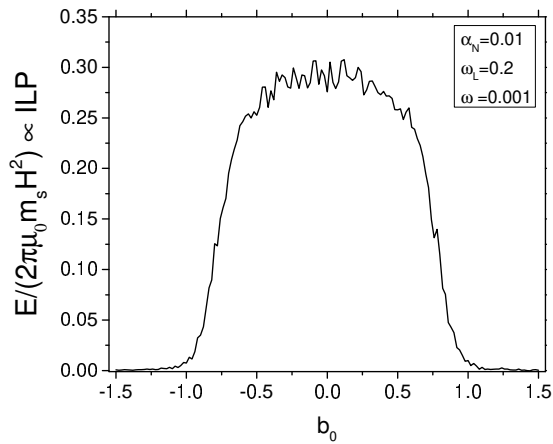}
\caption{Energy loss as a function of the {\it parallel} static field amplitude obtained by the 
stochastic LLG equation averaged over thirty four cycles with $\alpha_N = 0.01$, 
$\omega_L = 0.2$, $\omega = 0.001$ and $T = 300$K.
\label{fig9}
} 
\end{center}
\end{figure}

%%%%%%%%%%%%%%%%%%%%
As second step, we solve the stochastic LLG equation \eq{sLLG} in the presence of the 
applied magnetic field \eq{H_osc_parallel_def}, which stands for a parallel combination of static 
and oscillating fields. We plot in \fig{fig9} the energy loss as a function of the static field 
(for the the same set of parameters as before) for parallel orientations.
%
% Figure 9
%

Similarly to the case of vanishing static field, again thermal effects change drastically 
the dependence of the energy loss on the static field $b_0$ as seen if one compares 
\fig{fig4} and \fig{fig9}. The deterministic case predicts no energy loss for any 
non-vanishing value for $b_0$ (in the limit of $t\to \infty$), but the stochastic result 
changes this picture for relatively small values of $b_0$. The reason for
such significant modification is again the absence of time-dependent steady state 
solutions. Of course, large enough static field keeps aligned the magnetic moments, 
which results in a vanishing energy loss anyway.

%%%%%%%%%%%%%%%%%%%%
\subsection{Perpendicular static field, oscillating case}
\begin{figure}[ht] 
\begin{center} 
\includegraphics[width=7.2cm]{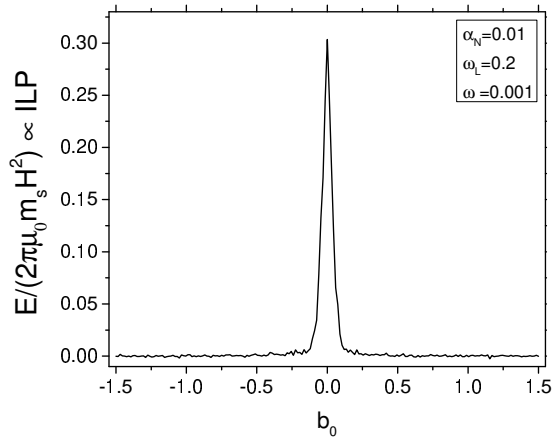}
\caption{Energy loss as a function of the {\it perpendicular} static field amplitude obtained by the 
stochastic LLG equation averaged over fourteen cycles with $\alpha_N = 0.01$, $\omega_L = 0.2$, 
$\omega = 0.001$ and $T = 300$K.
\label{fig10}
} 
\end{center}
\end{figure}

%%%%%%%%%%%%%%%%%%%%
We solve now the stochastic LLG equation \eq{sLLG} in the presence of the applied 
magnetic field \eq{H_osc_perp_def}, which stands for a perpendicular combination of static and 
oscillating fields. The energy loss as a function of the static field is plotted for perpendicular 
orientations in \fig{fig10}.
%
% Figure 10
%
In this case, thermal fluctuations only slightly modify the deterministic results.
The stochastic description confirms that with static field oscillating applied fields 
are not favoured, i.e., the largest energy loss can be achieved by the vanishing static field.

%--------------------------------------------------------------------------------------
\subsection{Discussion}
\label{sec_disc}
%--------------------------------------------------------------------------------------
In summary, thermal effects are found to be important for those cases only where 
the deterministic LLG equation has {\it no} time-dependent steady state solutions. 
These are the cases of vanishing and parallel static field configuration for oscillation. 
At variance, thermal fluctuations have no significant effects if the static field is in the 
plane of rotation or perpendicular to the line of oscillation, a case in which the 
deterministic (and stochastic) LLG equation has time-dependent steady state solutions. 
Thus, thermal effects have an important impact on superlocalization. We find 
that based on the results obtained from the stochastic LLG, the superlocalization 
effect of the deterministic case for oscillating fields in presence of a parallel static 
field does not hold, while it is conserved in the other considered cases: rotating fields 
with an in plane static field and oscillating field with a perpendicular static field.

%--------------------------------------------------------------------------------------
\section{Heating efficiency of rotating and oscillating cases}
\label{sec_compare}
%--------------------------------------------------------------------------------------
Let us compare the maximum heating efficiency or more precisely the ILP and SAR values
of the rotating and oscillating cases paying attention on the superlocalization effect too.
In order to achieve maximum energy loss for the rotating case, the static field should be 
in the plane of rotation and its amplitude should be identical to that of the rotating field (for
isotropic, i.e., spherically symmetric nanoparticles). For the oscillating field, the static field 
should vanish to achieve the maximum energy loss. 

%%%%%%%%
\subsection{Variable frequency and fixed amplitude for the applied field}
%%%%%%%%01

We can compare four cases: results obtained for oscillating applied fields with 
vanishing static field for $T=0$ and $T\neq 0$; and findings for the rotating case
with in-plane static field (with $\vert b_0 \vert =1$) for $T=0$ and $T\neq 0$. 

\begin{figure}[ht] 
\begin{center} 
\includegraphics[width=7.2cm]{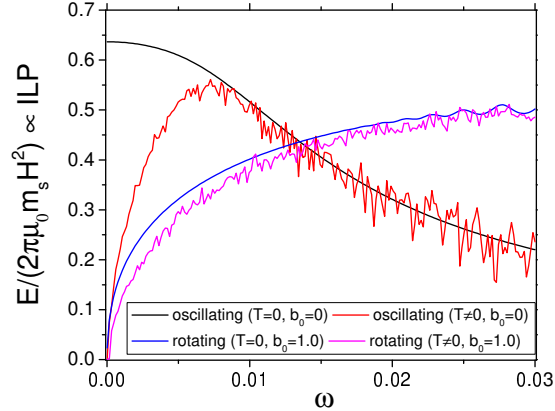}
\caption{(Color online.) Dependence of the energy loss/cycle on the applied (dimensionless) 
frequency averaged over eightyfour cycles with parameters $\alpha=0.1$,
$\alpha_N=0.02, \omega_L=0.2$ and $T=300$K.
\label{fig11}} 
\end{center}
\end{figure}

\begin{figure}[ht] 
\begin{center} 
\includegraphics[width=7.2cm]{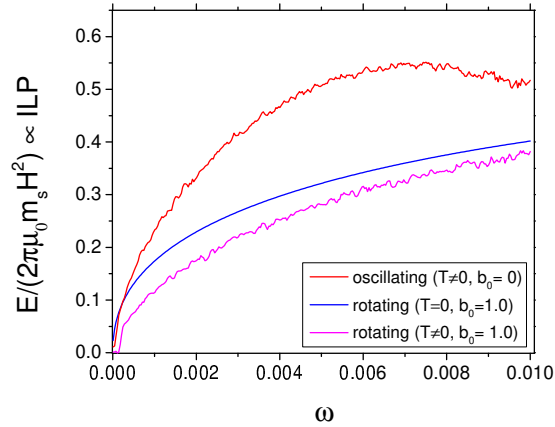}
\caption{(Color online.) Same as \fig{fig11}, but for the low-frequency limit.
\label{fig12}} 
\end{center}
\end{figure}

The dependence of the energy loss/cycle on the applied (dimensionless) frequency 
is plotted for the above mentioned four cases in \fig{fig11}. We remind that the 
dimensionless parameters are defined by \eq{dimless_param} and a typical value 
for the dimensionless frequency suitable for hyperthermia is $\omega \sim 10^{-4}$.
It is clear from \fig{fig11} that for the application of hyperthermia one should 
consider the low frequency limit of the four curves, which is plotted in \fig{fig12}.
%
% Figure 11
%
%
% Figure 12
%

We observe that a similar figure can be found in \cite{Raikher_Stephanov}, see Fig.~6 
therein where oscillating and rotating cases are compared, but without any static field. 
In addition, to check the stochastic solver, we compare our numerical data obtained
for the pure oscillating case (with no static field) to existing theoretical and experimental
results in \app{sec_check_results}. 

Static field is of great importance for the rotating case since in-plane static field can 
give raise to a drastic enhancement of the heating efficiency. Therefore the conclusions 
of \cite{Raikher_Stephanov} has to be examined in presence of a (possibly small) 
in-plane static field, a re-examination which is done here.

For the oscillating case (in the absence of static fields) the stochastic results
substantially modify for very low frequencies the picture emerging from the 
deterministic LLG. In the previous section we discussed in detail the reason 
for that and the explanation is related to the absence of time-dependent steady 
state motions, which results in an artifact for the limit of $\omega \to 0$ of the 
deterministic case. In other words in case of the absence of (time-dependent) 
steady state motions, the athermal and low-frequency limits do not commute.
This general statement is considered for the special case of oscillating external 
applied field in \cite{Raikher_Stephanov}. Thus, for oscillation (without static field) 
it is very important to rely on the stochastic results.

For the rotating case with in-plane static field, thermal effects are at variance
less important because the presence of attractive steady state motions. Thus, in 
this case one can rely on the deterministic results, however, for very 
low-frequencies the precise quantitative analysis requires the stochastic approach 
as shown in \fig{fig12}. The reason why the deterministic and the stochastic results 
are quantitatively different for the rotating case in the low-frequency limit is the 
following. If the magnitudes of the static and the rotating fields are identical which 
is the case for $\vert b_0 \vert =1$ then the effective applied field varying 
in time drops to zero when the static and rotating fields have opposite directions 
and this happens once in each and every cycle. At this point the magnetisation 
vectors of the individual nanoparticles feel no external field and they start to diverge
from each other due to thermal fluctuations. This effect is enhanced for low frequencies
when the variation of the effective applied field in time is slow compared to thermal 
effects, so, the deterministic and the stochastic results start to differ from each other
quantitatively. If the magnitudes of the static and the rotating fields are {\it not} 
identical, i.e., for $\vert b_0 \vert \neq 1$, then the effective applied field never
drops to zero, thus, one finds no difference between the deterministic and the
stochastic results.

\begin{figure}[ht] 
\begin{center} 
\includegraphics[width=7.8cm]{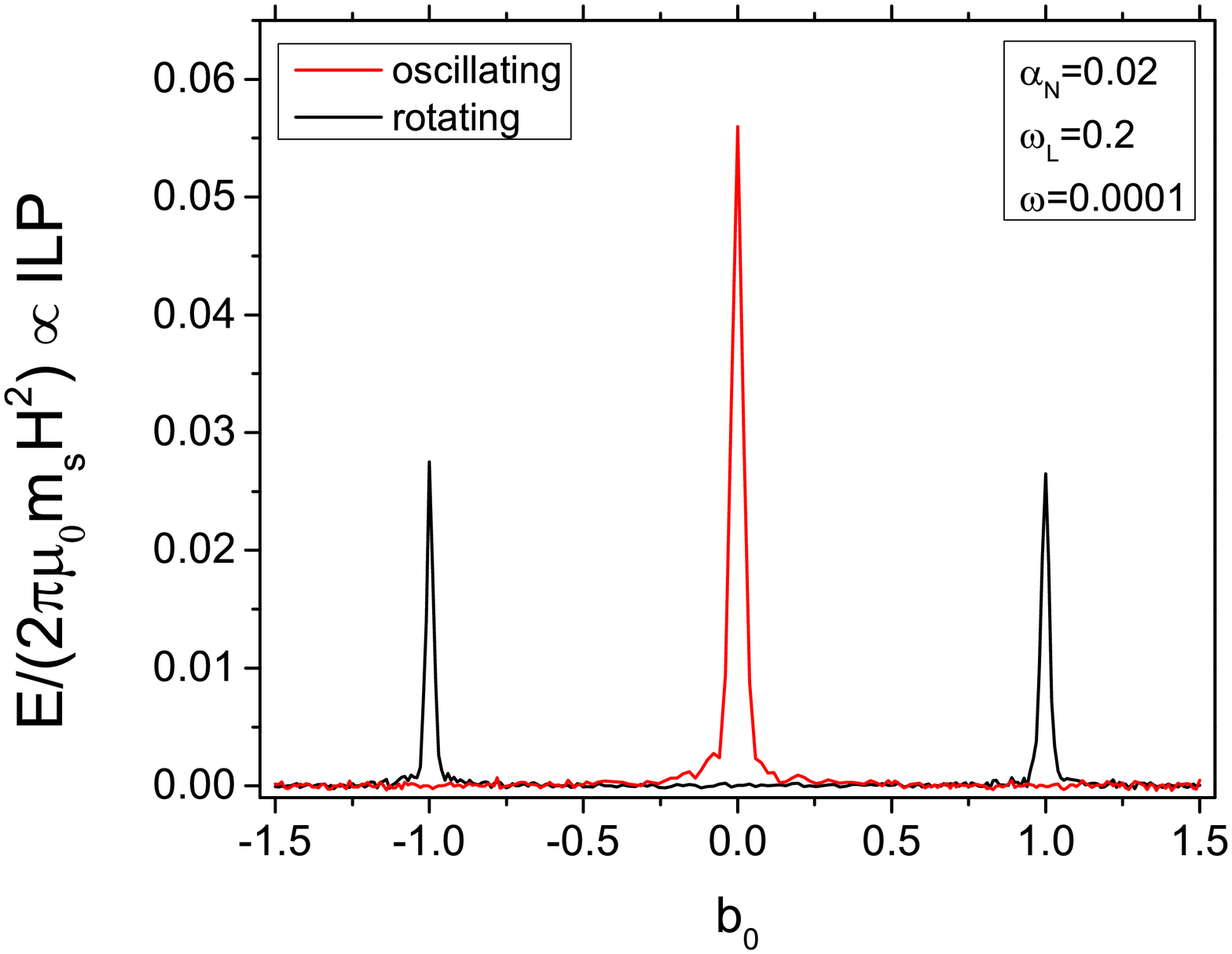}
\caption{(Color online.)
  Comparison of the energy loss as a function of the static field amplitude for 
the oscillating and the rotating cases averaged over twenty four cycles.
The static field is chosen to be perpendicular 
in the oscillating case and in-plane for the rotating applied field. The applied field
magnitude is $H = 18$kA/m, the applied frequency is chosen to be its maximum 
value $2 \times 10^{6}$Hz, the dimensionless damping parameter is $\alpha = 0.1$ 
and the temperature is $T = 300$K.
\label{fig13}} 
\end{center}
\end{figure}

The rotating case produces a better heating efficiency for very large frequencies, but these
are {\it not} suitable for hyperthermia. For smaller frequencies, the oscillating case becomes
more favourable. For very low frequencies, see \fig{fig13},
%
% Figure 13
%
the oscillating field for vanishing static field produces twice as much heating efficiency 
as that of the rotating one with in-plane static field with $\vert b_0 \vert=1$. The reason
is the following. We have discussed already that the effective applied field of
the rotating case drops to zero when the static and rotating fields have opposite directions
(and they have same magnitudes) and this happens once in each and every cycle.
One can also show that the majority of the energy loss over a single cycle is observed
when the effective field starts to grow up from zero. This is true for oscillating case, too.
Thus, when the rotating field is combined with an in-plane static field with identical 
amplitudes, the resulting effective applied field acts very much like an oscillating one 
in the very low-frequency limit. The difference between the two cases is the fact that 
for the oscillating case the effective field drops to zero two times per cycle, thus, its 
heating efficiency is expected to be twice as much large as the combined rotating one. 
Indeed, this can be seen, if one compares the peaks in \fig{fig13}. 
However, it is not obvious that the oscillating case remains favourable for a different
choice of the applied frequency and field strength keeping their product to be constant
to support the Hergt-Dutz limit.  Let us examine this in the next subsection.

%%%%%%%%
\subsection{Variable frequency and amplitude for the applied field}
%%%%%%%%

\begin{figure}[ht] 
\begin{center} 
\includegraphics[width=8.2cm]{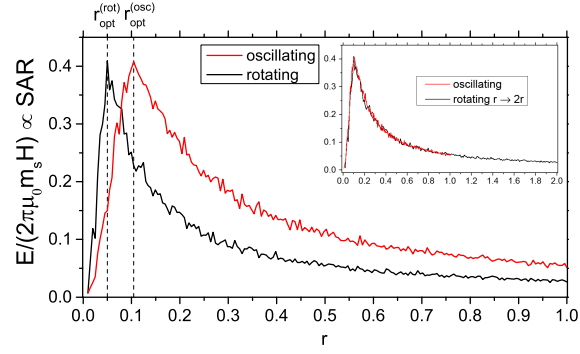}
\caption{(Color online.) Comparison of the SAR as a function of the ratio $r$ given 
by \eq{ratio_r} obtained for the oscillating (zero static field) and rotating (in-plane static field) 
cases averaged over fourteen cycles. One finds a single peak at $r^{\mr{(rot)}}_{\mr{opt}}= 0.05$ 
for the rotating case and at $r^{\mr{(osc)}}_{\mr{opt}}= 2\,\, r^{\mr{(rot)}}_{\mr{opt}} = 0.1$ 
for the oscillating one. The inset shows that the two curves coincide if the SAR of the
rotating case is plotted as a function of $2 r$.
\label{fig14}}
\end{center}
\end{figure}

It is shown in \fig{fig13} that the appropriate combination of rotating and (in-plane) 
static fields acts like an oscillating one (for relatively low frequencies) but in order to 
achieve the same heating efficiency or more precisely the same SAR one has to double 
the applied frequency. Thus, if one plots the SAR as a function of the ratio $r$ of the 
applied field strength (amplitude) $H$ and the angular frequency $\omega$ and keeping 
their product $H \omega = H_\star \omega_\star$ at a constant value,
\beq
\label{ratio_r}
r^2 = \frac{H}{\omega} \frac{\omega_\star}{H_\star}
\hskip 0.3cm \to \hskip 0.3cm
r^2 \equiv 1 \,\,\, \mr{for} \,\,\, H_\star = 18 \mr{kA/m}, \,\,\, \omega_\star = 2 \times 10^6 \mr{Hz}
\hskip 0.3cm \to \hskip 0.3cm
r =  \sqrt{\frac{\omega^2_\star}{\omega^2}} = \sqrt{\frac{H^2}{H^2_\star}}
\eeq 
one expects the same type of function for the oscillating and rotating cases but 
shifted by a factor of two, see \fig{fig14}.
%
% Figure 14
%
Indeed, on \fig{fig14} we show the SAR as a function of the ratio $r$ and one finds a 
single maximum for the rotating and oscillating cases where the positions of the peaks 
are related to each other, $r^{\mr{(osc)}}_{\mr{opt}}= 2\,\, r^{\mr{(rot)}}_{\mr{opt}}$.
This demonstrates that in order to achieve the same heating efficiency, i.e., the same 
SAR for the rotating and oscillating cases, one has to {\it double} the applied frequency
(more precisely the ratio $r$) of the rotating applied field compared to the oscillating 
one, so, in this case the SAR functions coincide, see the inset of \fig{fig14}. 

\begin{figure}[ht] 
\begin{center} 
\includegraphics[width=7.0cm]{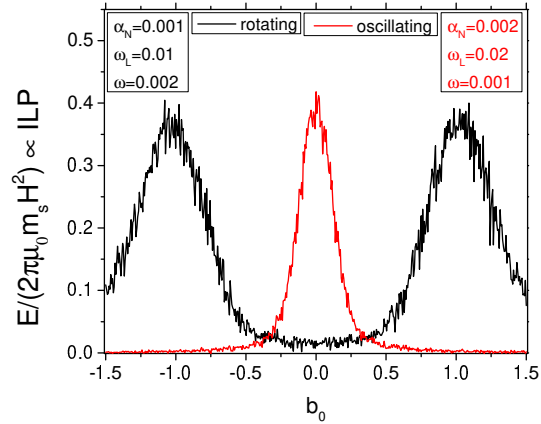}
\caption{(Color online.) In this figure we show how the superlocalization effect presented 
on \fig{fig13} is modified for smaller frequencies and larger field strength values averaged 
over thirty four cycles. We choose the following dimensionful parameters for the 
oscillating $2 \times 10^7$Hz, $1800$A/m and for the rotating $4 \times 10^7$Hz, 
$900$A/m cases. 
\label{fig15}} 
\end{center}
\end{figure}

The peaks suggest optimal choices for the applied frequency and field 
strength (amplitude) for the oscillating ($2 \times 10^7$Hz, $1800$A/m) and for the 
rotating ($4 \times 10^7$Hz, $900$A/m) cases. At these values, the SAR's and ILP's are the 
same, see \fig{fig15}.
%
% Figure 15
%
One finds identical SAR and ILP values for the rotating and the oscillating cases but at 
different frequencies and different field amplitudes. In particular, for smaller field 
amplitudes and for higher frequencies the combination of the rotating and in-plane 
static fields seems to be a better choice, i.e., it gives larger SAR. 
However, we have two important comments. First, the superlocalization effect 
is weakened compared to the case discussed in the previous subsection where 
the frequency was smaller ($2 \times 10^6$Hz), and the field strength was larger 
($18$kA/m). Second, one has to keep the applied frequency below the the eddy 
current threshold, so the maximum applied frequency cannot exceed 1000 kHz 
(under any circumstances), thus, the peaks of the SAR graphs cannot be used in 
medical applications.

\begin{figure}[ht] 
\begin{center} 
\includegraphics[width=8.2cm]{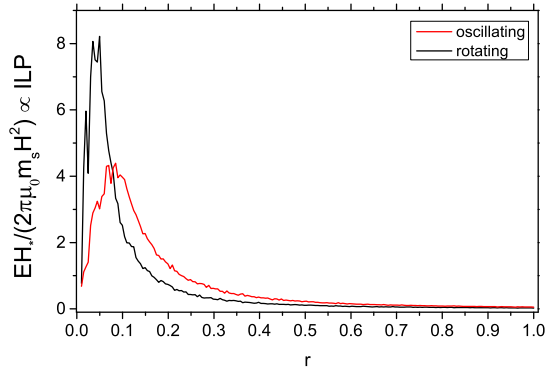}
\caption{(Color online.) The same as \fig{fig14}, but for the ILP values.
\label{fig16}} 
\end{center}
\end{figure}

Instead of the SAR, it is more reliable to study the ILP values,
see, \fig{fig16} which has huge importance in technical realisation. 
%
% Figure 16
%
The ILP values of \fig{fig16} show that in the low frequency (large $r$) 
regime suitable for hyperthermia, the oscillating applied field is shown to result in two 
times larger ILP then the rotating one with identical superlocalization ability which 
has importance in technical realisation.

Let us note that we expect that the above findings do not depend on the maximum choice of
the product of the frequency and field strength $\omega_\star H_\star$. 
This is because the deterministic LLG equation is invariant under the appropriate simultaneous 
rescaling of $\omega$, $H$ and the time variable $t$, thus these quantities 
are related to each other, so one can draw some conclusions on from 
\fig{fig14} and \fig{fig16} for other values of the product $\omega_\star H_\star$.

%--------------------------------------------------------------------------------------
\section{Summary and Conclusions}
\label{sec_sum}
%--------------------------------------------------------------------------------------
In this work we considered theoretically  whether and how thermal 
effects can modify the enhancement of heat transfer and the superlocalization effect 
of magnetic nanoparticle hyperthermia for oscillating and rotating applied fields.
Furthermore, we compared the efficiency and superlocalization of the rotating 
and oscillating cases.

It is known that if the amplitudes of the (in-plane) static and rotating applied fields 
fulfil a certain relation (they should be identical for the isotropic case), then the 
energy transfer from the applied field into the environment is found to be 
enhanced drastically. In addition, by using an inhomogeneous static field
one can induce superlocalization. Another example for superlocalization is the 
combination of an applied oscillating and a static field gradient, which provides 
a spatially focused heating since for large enough static field the dissipation drops 
to zero, so, the temperature increase is observed only where the static field vanishes. 

It is a relevant question to ask whether thermal effects modify the sharpness of 
superlocalization. Therefore for experiments, i.e., application in tumor-therapy, 
thermal fluctuations have to be taken into account. Thus, it is required to study 
these issues by the stochastic  Landau-Lifshitz-Gilbert (LLG) equation, whose 
investigation is presented here.

We obtained that when time-dependent steady state motions are 
present, then deterministic and stochastic results are very similar to each other. 
This is found in the rotating case with in-plane static field and in the oscillating case 
with perpendicular static field. However, in the {\it absence} of these time-dependent  
steady state motions, the efficiency of the method and its superlocalization suffers 
drastic modifications by thermal effects. This occurs for the oscillating case with 
vanishing and parallel static fields.

Our most important result is that comparing the most efficient rotating and 
oscillating combinations, we found that for smaller field amplitudes and for higher 
frequencies the combination of the rotating and in-plane static fields seems to be 
a better choice, i.e., it gives larger ILP and SAR values however, 
these high frequencies are unacceptable for medical applications. In the 
low-frequency range suitable for magnetic hyperthermia, the oscillating applied
field was shown to result in two times larger ILP and SAR then the rotating one 
with identical superlocalization ability which has importance in technical realisation.

In particular, we showed that one finds identical SAR values for the rotating and the 
oscillating cases but at different frequencies and different field amplitudes. In other
words, if one compares the rotating and oscillating applied fields at the same 
frequency suitable for medical applications, the latter produces us a better heating 
efficiency. This theoretical result holds with only one condition, namely that the 
energy loss over the cycle must be localised to the time-domain when the applied 
field starts to increase from its zero value, so thus the appropriate combination of 
rotating and (in-plane) static fields acts like an oscillating one.

In summary, we demonstrated that the combination of static and rotating 
fields can have importance in technical realisation of magnetic nanoparticle 
hyperthermia but keeping in mind that it has a worse heating efficiency compared 
to the usual oscillating one in the frequency range suitable for medical applications.

Of course, in order to look for the most efficient applied field configuration, one can 
perform further theoretical investigations. For example, open questions motivated by 
the present theoretical study of the enhancement and superlocalization effects are 
the inclusion of {\it (i)} the interaction between the particles which can results in 
formation of clusters \cite{usov_claster}; {\it (ii)} the dependence on the diameter of the 
nanoparticles; and {\it (iii)} the possible rotating motions of the nanoparticle as a whole 
\cite{lyutyy_general,joined_motion,joined_motion_bis_1,joined_motion_bis_2,
joined_motion_2,thermal_and_dipole,viscous_rotating}.
However, we do expect that considerations very likely do not violate our major finding 
that appropriate combination of rotating and (in-plane) static fields is essentially like an 
oscillating one but in order to achieve the same heating efficiency, i.e., the same SAR 
one has to double the applied frequency, so, for the same frequency, 
the oscillating combination is a better choice.

Based on our results, we suggest targeted experimental studies incorporating
standardized techniques \cite{standardized_techniques,wildeboer_r_r}.  For example, we 
propose to use well separated isotropic nanoparticles in an aerogel matrix where 
only the magnetic moment can rotate. There are proposals for the technical realisation 
of a rotating applied field, see, e.g.,~\cite{exp_rot} and \cite{non_calorimetric}.  
For example, in \cite{exp_rot} the authors discuss possible experimental realisation 
of generating a rotating applied field. The standard choice could be a system of two pairs 
of Helmholtz coils whose axes are perpendicular to each other but in \cite{exp_rot} the
authors suggest a system of three pairs of inductors connected in series with capacitors 
to generate a rotary magnetic field. This configuration (with the inclusion of a static field) 
can be directly used to test theoretical predictions of the present work. 
Also non-calorimetric technics \cite{highly_accurate}, \cite{non_exp_relax} are used to 
measure the energy losses in various applied fields.  
In \cite{non_calorimetric}, the authors present the validation of a resonators with a 
so-called birdcage coil. This represents another realisation of rotating magnetic field.
Again, with the inclusion of a static field on can test our numerical results obtained
for the rotating field case. In addition, one can discuss possible experimental 
realisation to measure the decrease of heat generation as a function of the static field. 
For example, a possible measurement setup could be when a rotatable resonator is 
placed into the electromagnet which generates the static field. In this manner, one
can easily perform measurements for the case when the oscillating filed and the 
static one are perpendicular or parallel to each other.
Thus, the predictions of the present work can be tested with the help of the required 
combinations of static and alternating applied fields.

\section*{Acknowledgement}
The CNR/MTA Italy-Hungary 2019-2021 Joint Project "Strongly interacting systems in 
confined geometries" is gratefully acknowledged. Partially supported by the \'UNKP-20-4-I 
New National Excellence Program of the Ministry for Innovation and Technology from the 
source of the National Research, Development and Innovation Fund. Fruitful discussions 
with Giacomo Gori and Stefano Ruffo are gratefully acknowledged.

\appendix
%--------------------------------------------------------------------------------------
\section{Consistency checks of the stochastic LLG equation}
\label{sec_checks}
\begin{figure}[ht] 
\begin{center} 
\includegraphics[width=7.8cm]{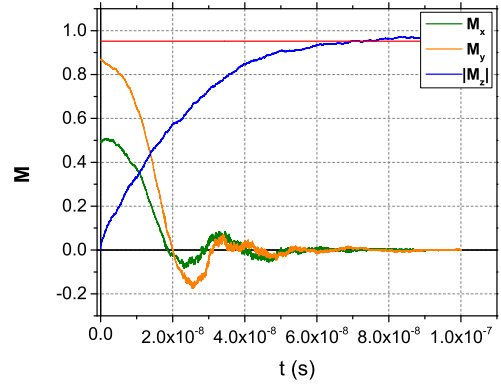}
\caption{(Color online.) In this figure we plot the average values of $M_x(t)$ 
(green curve), $M_y(t)$ (yellow curve) and $M_z(t)$ (blue curve) are plotted versus the 
time where the horizontal red line stands for the asymptotic value given in Ref.~\cite{Giordano}
(compare it in particular with the central panel of Fig. 6 in Ref.~\cite{Giordano}).
\label{fig17}
} 
\end{center}
\end{figure}

%--------------------------------------------------------------------------------------
In this Appendix we present some consistency checks based on the numerical results
discussed \cite{Giordano}. In section VI of \cite{Giordano} the magnetization dynamics 
of an anisotropic nanoparticle has been studied in the absence of any applied field and
the anisotropy field is assumed to be parallel to the z-axis (by using uniaxial shape anisotropy):
\bea
\label{H_giordano}
{\bf H}_{\rm{eff}} = {\bf H}_{\rm{a}} = H \, \, (0, 0, \lambda_{\rm{eff}} M_z),
\eea
where $\lambda_{\rm{eff}} = H_a /H$ is the anisotropy ratio and $H_a$ is the 
anisotropy field.
In the spherical coordinate system the stochastic LLG equation \eq{sLLG} takes the form 
\begin{align}
\frac{\rm{d}}{\rm{d}t} \phi &= 
-\omega_L \lambda_{\rm{eff}} \cos \theta +
\frac{1}{\sin\theta} \sqrt{\frac{1}{2\tau_N}} n_\phi,
\\
\frac{\rm{d}}{\rm{d}t} \theta &= 
-\alpha_N \lambda_{\rm{eff}} \cos \theta \sin \theta+
\frac{1}{2\tau_N} \frac{\cos\theta}{\sin\theta} + 
\sqrt{\frac{1}{2\tau_N}} n_\theta,
\end{align}
where $n_\phi$ and $n_\theta$ satisfy the relations given in \eq{nproperties}.
The above stochastic differential equations are identical to Eq.~(27) of \cite{Giordano}. 
To have a more clear picture, these can be rewritten to the standard form of an Ito process 
$\mathrm{d} X = b(X) \, \mathrm{d} t + \sigma \, \mathrm{d} W$. 
From \eq{nproperties} one can observe that the variance of $n_{\phi(\theta)}$ is multiplied 
by a factor of 2, which can be explicitly written as a $\sqrt{2}$ multiplicative factor of the 
Wiener process. Thus the full stochastic process in the standard form writes as
\begin{align}
\rm{d} \phi &= 
-\omega_L \lambda_{\rm{eff}} \cos \theta \; \rm{d}t +
\frac{1}{\sin\theta} \sqrt{\frac{1}{2\tau_N}} \sqrt{2}\; \rm{d}W_1,
\\
\rm{d} \theta &= 
\left(-\alpha_N \lambda_{\rm{eff}} \cos \theta \sin \theta+
\frac{1}{2\tau_N} \frac{\cos\theta}{\sin\theta} \right) \rm{d}t + 
\sqrt{\frac{1}{2\tau_N}} \sqrt{2}\; \rm{d}W_2,
\end{align}
where $\rm{d}W_1$ and $\rm{d}W_2$ are independent Wiener processes with the standard normal distribution.
This Ito process can be directly solved by the built-in functions of Mathematica \cite{wolfram}, where the time is 
discretized. In this case setting the numerical time step ${\rm{d}t}=1.5\times 10^{-10}s$ produces reliable results. 
At each time step, $\rm{d} \phi$ and $\rm{d} \theta$ receives a random, normal distributed 'kick' from the $\rm{d}W$ 
terms centered around zero with a variance that is adjusted by the size of the time step $\sigma^2_W= \rm{d}t$.
Fig.~6 of Ref.~\cite{Giordano} contains some details regarding the average values of the 
Cartesian components of the magnetization vector. Therefore it is illustrative to recover 
this known result by our stochastic solver using the same values for the parameters
$\tau_N=1.5 \times 10^{-7}$s, $\omega_L \lambda_{\rm{eff}}=2.58\times 10^8$s$^{-1}$, 
$\alpha_N \lambda_{\rm{eff}}=7.74\times 10^7$s$^{-1}$, results from which are presented in \fig{fig17}.
The comparison confirms the validity of the numerical method used here for the 
stochastic solver.
%
% Figure 17
%
%

%--------------------------------------------------------------------------------------
\section{Consistency check of the results}
\label{sec_check_results}
%--------------------------------------------------------------------------------------
In order to check our theoretical results based on the stochastic LLG equation,
we determine the imaginary part of the AC susceptibility obtained for the pure
oscillating (without static) external field and compare it to the corresponding 
literature both on the theory and the experimental side. The frequency 
dependence of the AC susceptibility is given by
\bea
\label{chi_imaginary}
\chi''(\omega) =  \chi_0 \frac{\omega \tau}{1+\omega^2 \tau^2} 
\eea
see for example Eq.(4) and (5) of \cite{non_exp_relax}. This form is valid when
a single relaxation process is present only. In our case the frequency dependence 
is characterised by the N\'eel relaxation time, $\tau = \tau_N$, since we investigated
small enough nanoparticles to neglect their rotation inside their environment and 
only the N\'eel process, i.e, the rotation of the magnetisation is taken into count.

\begin{figure}[ht] 
\begin{center} 
\includegraphics[width=7.8cm]{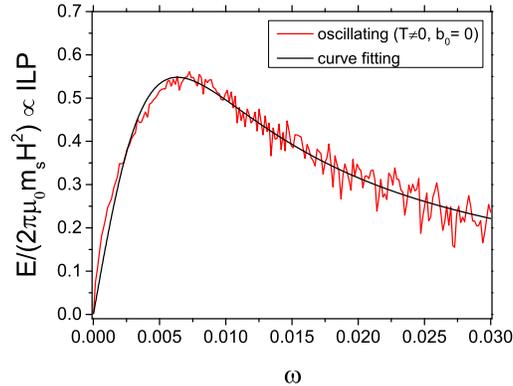}
\caption{(Color online.) In this figure we plot the imaginary parts of the AC susceptibility 
(which is related to ILP) as a function of the applied (dimensionless) frequency for pure 
oscillating case. Numerical data are identical to the red curve of \fig{fig11}. The solid 
line is a fitted curve to numerical data based on \eq{chi_imaginary}.
\label{fig18}
} 
\end{center}
\end{figure}

We use the formula \eq{chi_imaginary} to fit our theoretical results obtained by 
the stochastic LLG equation for the pure oscillating case. On \fig{fig18} the solid 
line denotes the fitted curve which gives the following fitted relaxation time,
$\tau_{N} = 157 \times 0.5 \times 10^{-10} \mr{s} \sim 0.78 \times 10^{-8} \mr{s}$
and the following fitted magnitude, $\chi_0 = 1.09$. 
%
% Figure 18
%

We conclude that our numerical results obtained by the stochastic LLG equation 
can be well described by the standard formula \eq{chi_imaginary}. In addition, 
the relaxation time calculated by the fitting procedure, $\tau_N \sim 1$ ns is 
in the range of typical value for magnetic nanoparticles, see e.g., \cite{ota}. 
Finally, we argue that our results are in agreement with known experimental 
results, see for example, Fig.2 of \cite{ferguson}.

\end{document}